\documentclass[preprint,showpacs,preprintnumbers,amsmath,amssymb]{revtex4}
\usepackage{epsfig,amsmath,amssymb,graphics,color,calc,wasysym}

\newcommand{\be}{\begin{equation}}
\newcommand{\ee}{\end{equation}}
\newcommand{\ba}{\begin{eqnarray}}
\newcommand{\ea}{\end{eqnarray}}

\renewcommand{\phi}{\varphi}

\begin{document}

\title{Dynamical Heterogeneities and Cooperative Motion in Smectic Liquid Crystals}

\author{Alessandro Patti}
\affiliation{Soft Condensed Matter, Debye Institute for NanoMaterials Science, Utrecht University, Princetonplein 5, 3584 CC, Utrecht, The Netherlands}

\author{Djamel El Masri}
\affiliation{Soft Condensed Matter, Debye Institute for NanoMaterials Science, Utrecht University, Princetonplein 5, 3584 CC, Utrecht, The Netherlands}

\author{Ren\'{e} van Roij}
\affiliation{Institute for Theoretical Physics, Utrecht University, Leuvenlaan 4, 3584 CE, Utrecht, The Netherlands}

\author{Marjolein Dijkstra}
\affiliation{Soft Condensed Matter, Debye Institute for NanoMaterials Science, Utrecht University, Princetonplein 5, 3584 CC, Utrecht, The Netherlands}

\date{\today}

\begin{abstract}
Using simulations of hard rods in smectic-A states, we find
non-gaussian diffusion and heterogeneous dynamics due to the {\em
equilibrium} periodic smectic density profiles, which give rise to
permanent barriers for layer-to-layer diffusion. This relaxation
behavior is surprisingly similar to that of {\em non-equilibrium}
supercooled liquids, although there the particles are trapped in
transient (instead of permanent) cages. Interestingly, we also
find stringlike clusters of up to 10 inter-layer rods exhibiting
dynamic cooperativity in this equilibrium state.
\end{abstract}


\maketitle

Smectic liquid crystals consist of stacks of fluidlike layers of
oriented rodlike particles \cite{degennes}. They can be stabilized
by attractive Van der Waals forces, e.g. in thermotropic systems
of mesogenic molecules \cite{degennes}, or by repulsive hard-core
interactions between sufficiently elongated particles
\cite{frenkel}. In the relatively simple smectic-A phase the
static structure is characterized by long-range orientational
ordering of the rods combined with a one-dimensional periodic
density variation in the direction parallel to the rods. While the
equilibrium properties of smectic phases are relatively well
understood \cite{degennes}, little is known about their dynamics
on the particle scale, even though Helfrich's early report of
diffusion ("permeation") of anisotropic particles through smectic
layers goes back 40 years \cite{helfrich}. Recently, however,
exciting progress was made based on newly developed experimental
techniques (e.g. NMR coupled to strong magnetic field gradients
\cite{furo}, or fluorescent labelling of rods \cite{lettinga}),
which revealed direct observations of non-Gaussian diffusion and
quasiquantized layer-to-layer hopping across a barrier
\cite{lettinga}. This triggered new theoretical work, based on
dynamic density functional theory (DDFT), which not only confirmed
the non-Gaussian diffusive motion and the one-dimensional
"permanent" barriers due to the static smectic background, but also
showed the importance of "temporary" cages formed by neighboring
rods \cite{bier}.

Interestingly, non-Gaussian diffusive behavior due to
heterogeneous dynamics of  "slow" and "fast" particles is also a
key feature in glassy systems and supercooled liquids, in which
individual particles are trapped in transient cages formed by
their neighbors. This heterogeneous dynamics explains the
experimentally observed nonexponential relaxation and stretching
of time correlation functions and is found to be closely related
to cage rearrangements and to cooperative motion, in which a small
fraction of the particles (typically a few percent) move
collectively in stringlike \cite{kob,donati} or compact \cite{appignanesi, berthier} clusters .
Several models based on a heterogeneous distribution of diffusion
coefficients, jump times or jump lengths have been introduced to
explain the heterogeneous dynamics due to temporary cages \cite{chaudhuri}, while
another model explains the non-Gaussian diffusion by the dynamics
of a single Brownian particle in a periodic external potential,
which is very similar to the permanent barriers of the smectic-A
phase \cite{vorselaars}. The intriguing question that we address
in this Letter is to what extent the dynamics in the smectic-A
phase is heterogeneous and/or collective, or, in other words, to
what extent is the {\em equilibrium} smectic-A dynamics resemblant
to that of {\em out-of-equilibrium} quenched supercooled liquids?
This issue cannot be addressed directly in fluorescence
experiments, in which only a small fraction of the rods is
labelled such that moving clusters cannot be observed. While DDFT
yields average quantities such as density profiles, Van Hove
functions, and single-particle barriers, but does {\em not}
provide cluster information \cite{bier}. We therefore resort to
computer simulations of hard-rod fluids in the smectic phase.
Contrary to the essentially one-particle analysis of Ref.
\cite{cinacchi}, we do find clear evidence for heterogeneous
dynamics and cooperative permeation of strings of 1-10 rods.

We perform kinetic Monte Carlo (KMC) simulations in the
isobaric-isothermal ensemble $(NPT)$ of $N=1530-3000$ freely
rotating hard spherocylinders with aspect ratio $L^{*}\equiv
L/D=5$, where $D$ is the diameter and $L+D$ the length of the
rods. For $L^*=5$ the smectic bulk phase melts into a nematic
phase below $P^*=1.4$ and freezes into a crystal above $P^*=2.3$,
where $P^*=PD^3/k_BT$ is the reduced pressure with $k_B$
Boltzmann's constant \cite{bolhuis}. We study the dynamics at two
state points in the bulk smectic phase characterized by
$P_1^*=1.6$ and $P_2^*=2.0$, which correspond to packing fractions
$\eta_{1}=0.508$ and $\eta_{2}=0.557$, respectively. Translational
and rotational moves are attempted and accepted if no overlap is
detected. A rectangular simulation box with 5-10 smectic layers
and periodic boundary conditions are employed. Volume changes are
attempted every $N$ MC steps by randomly changing the side
length of the simulation box. In KMC simulations it is convenient
to relate the number of MC cycles to time $t$. We chose
$\tau\equiv D^{2}/D_{tr}$ as our unit of time, where $D_{tr}$ is
the translational short-time diffusion coefficient, which is the
isotropic average of the diffusion coefficients in the three space
dimensions.

Denoting the positions of the rods by ${\bf r}_i=(x_i,y_i,z_i)$,
we first measure the (relative) probability $\pi(z)$ of finding a
rod at position $z$, where the $z$-axis is parallel to the nematic
director $\hat{n}$. Following Ref. \cite{lettinga} we introduce
the Boltzmann factor $\pi(z)\propto \exp\left(-U(z)/k_{B}T\right)$
with $U(z)$ the effective potential for diffusion out of the
middle of a smectic layer. Fig. 1 shows $U(z)$ in a small
$z$-regime for our two state points together with the fit
$U(z)=U_{0}\sin^{2}\left(\pi z/h \right)$, with barrier height
$U_{0}=(3.5,7.5)k_BT$ and layer spacing $h=(L+D)(1.05,1.03)$ as
fit parameters for the two states. The denser state thus reveals a
substantially larger diffusion barrier and a stronger confinement
to the middle of the smectic layers.

\begin{figure}[!ht]
\center
\includegraphics[width=0.8\textwidth]{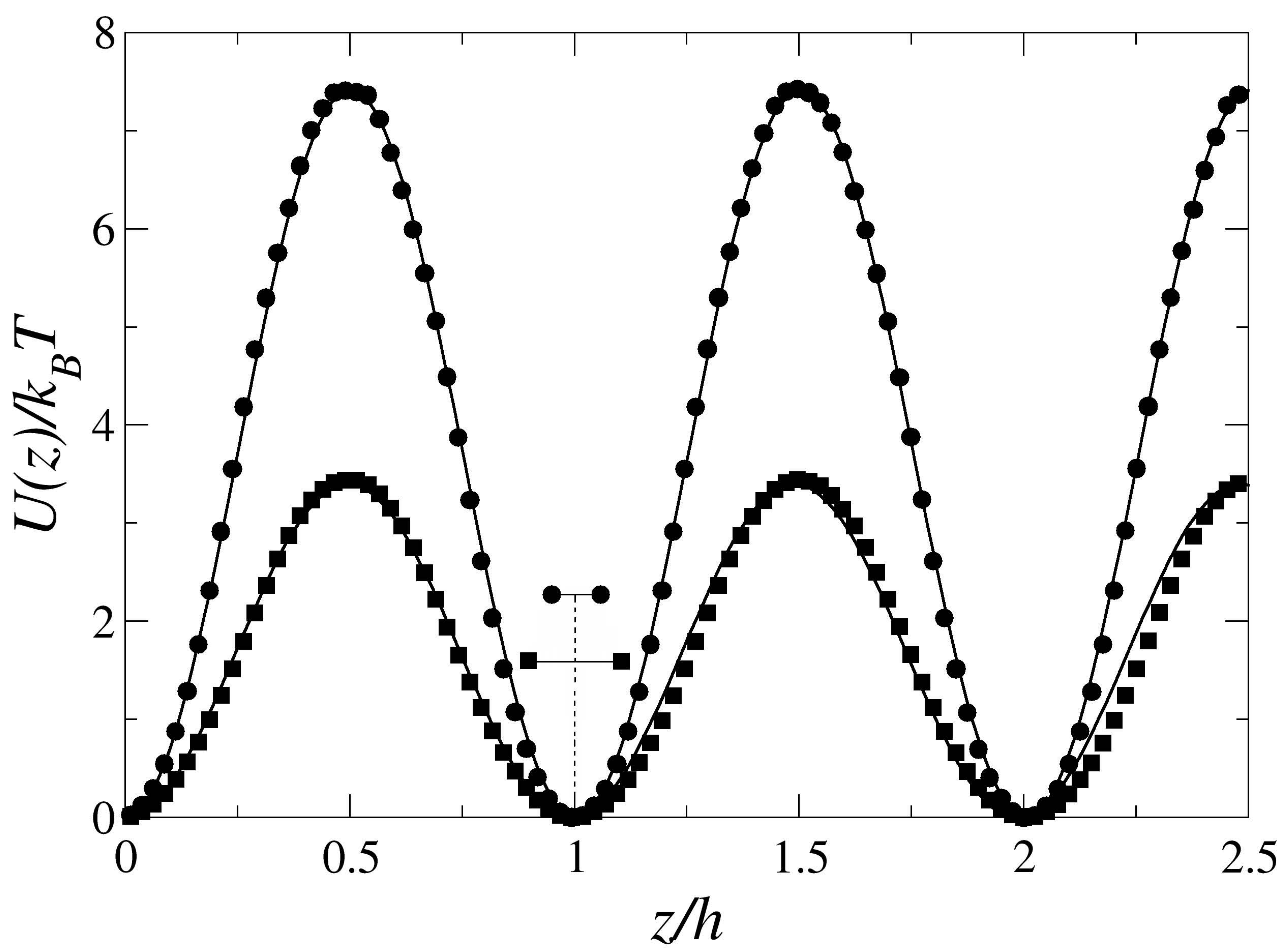}
\caption{Effective potential $U(z)$ at two state points in the bulk smectic phase with packing fractions
$\eta_1=0.508$ (red squares) and $\eta_2=0.557$ (black circles); the lines are fits. The horizontal bars denote
the standard deviation $\sigma=(0.6,0.3)D$ at $\eta_1$ and $\eta_2$, respectively.}
\end{figure}

In order to further quantify the physical picture of a
\textit{hopping-type} $z$-diffusion, with the rods rattling around
in a given layer until they overcome the free-energy barrier and
jump to a neighboring layer, we calculate the self part of the Van
Hove correlation function (VHF)
$G_{s}(\textit{z},t)=\frac{1}{N}\left\langle \sum_{i=1}^N
\delta\left[\textit{z}-(\textit{z}_{i}(t_0+t)-\textit{z}_{i}(t_0))\right]\right\rangle
$, with $\left\langle ... \right\rangle$ the ensemble average over
all particles and initial time $t_0$, and $\delta$ the
Dirac-delta. Note that $G_{s}(\textit{z},t)$ gives the
distribution for the particle $z$-displacements during a
time-interval $t$, and would be a Gaussian of $z$ for freely
diffusive particles. Fig. 2(a,b) show the VHF for our two state
points as a function of $z$ for several equidistant $t$'s, showing
the appearance of peaks at integer layer spacings consistent with
earlier experimental \cite{lettinga} and theoretical \cite{bier}
results. The height and spatial extension of the peaks is larger
at $\eta_{1}$, indicative of a faster layer-to-layer diffusion
than at $\eta_2$.
\begin{figure}[!ht]
\center
\includegraphics[width=0.8\textwidth]{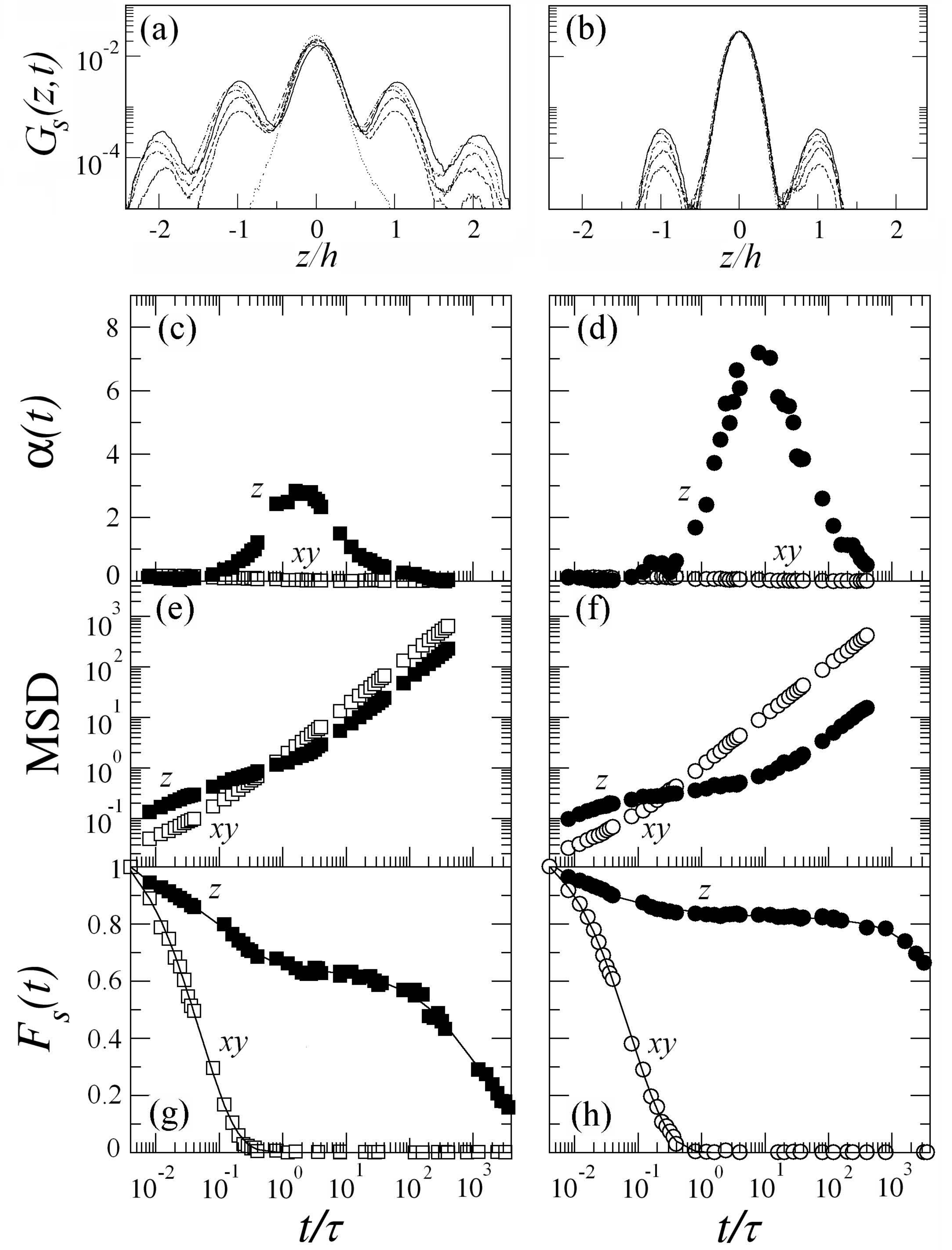}
\caption{(a,b) Self part of the Van Hove function $G_s(z,t)$,
(c,d) non-Gaussian parameters $\alpha_z(t)$ and $\alpha_{xy}(t)$,
(e,f) mean-square displacement (MSD) in units of $D^2$, and (g,h)
self-intermediate scattering functions $F_s(t)$, at packing
fraction $\eta_1=0.508$ (left column) and $\eta_2=0.557$ (right
column). The time intervals in (a,b) range from $t=0.4\tau$
(dotted lines) to $t=40\tau$ (solid lines), with increments of
$\approx8\tau$. The solid and open symbols refer to the $z$ and
$xy$ direction, respectively. }
\end{figure}
Non-gaussian VHF's were analyzed in 2D liquids \cite{hurley},
glasses \cite{chaudhuri}, and liquid crystals
\cite{lettinga,bier}, and attest to the presence of dynamical
heterogeneities which imply particles diffusing faster or slower
than the average. This is quantified by the well-known
non-gaussian parameter (NGP) \cite{rahman}, for $z$-diffusion
defined by $ \alpha_z(t)=-1+\langle \Delta
z(t)^{4}\rangle/3\langle \Delta z(t)^{2}\rangle^{2}$ with $\Delta
z(t)=z(t_{0}+t)-z(t_{0})$ the $z$-displacement of a rod in the
time interval $t$ starting at $t_0$. Likewise non-Gaussian
in-plane diffusion can be characterised by an NGP
$\alpha_{xy}(t)$. Heterogeneous dynamics occurs on a time-scale
$t$ if $\alpha_z(t)$ or $\alpha_{xy}(t)$ is non-vanishing. In Fig.
2(c,d) we show the NGP's for our two state-points, with
$\alpha_z(t)$ showing a clear peak at $t^*\simeq(2, 10)\tau$,
while $\alpha_{xy}(t)$ is hardly visible on the scale of the
figure and thus essentially vanishes for all $t$. In other words,
in-plane diffusion is dynamically homogeneous while permeation is
heterogeneous, with $t^*$ and $\alpha_z(t^*)$ increasing with
density. In order to determine the relaxation dynamics we plot the
mean square displacements (MSDs) $\langle\Delta z^2(t)\rangle$ and
$\langle\Delta x^{2}(t)+\Delta y^2(t)\rangle$ in Fig. 2(e,f) for
the two state points. The $xy$-MSD shows a smooth crossover from
short- to long time diffusion, while the $z$-MSD develops an
intermediate cage-trapping plateau up to $t^*$ beyond which
interlayer diffusion takes place. Note that the $z$-MSD exceeds
the $xy$-MSD only at short times.

In addition we study structural relaxation by computing the
self-intermediate scattering function
$F_{s}(t)=\langle\exp[i\textbf{q}\cdot\Delta
\textbf{r}(t)]\rangle$ at wavevectors ${\bf q}D=(0,0,q_{z})$ and
$(q_{x},q_{y},0)$, with $q_{z}=1$ and $(q^{2}_{x}+q^{2}_{y})^{1/2}=6$, 
that correspond to the main peaks in the static
structure factor. Here $\Delta {\bf r}(t)$ denotes the particle
displacement during a time-interval $t$.  Results are shown in
Fig. 2(g,h). The in-layer dynamics is very fast with nearly
exponential decay, typical of fluid-like behavior. By contrast,
the inter-layer dynamics is much slower, with $F_{s}(t)$ decaying
in two steps at $t'$ and $t''$ separated by a plateau during
$t'<t<t''$. Interestingly  this plateau (corresponding to
$\beta$-relaxation) coincides with the nonvanishing of
$\alpha_z(t)$, and can hence be attributed to the heterogeneity of
the dynamics \cite{kob}. The first step at $t'/\tau\leq 1$
corresponds to the rattling of rods inside the temporary cage and
permanent smectic background formed by neighboring rods
\cite{bier}, whereas the second one ($\alpha$-relaxation)
corresponds to the escape on a time scale that increases from
$t''/\tau\simeq 100$ at $\eta_1$ to $10^3$ at $\eta_2$. The
increase of $t''$ and that of the height of the plateau of
$F_s(t)$ with density is also observed for colloidal glasses
\cite{elmasri}. At $\eta_{1}$, the long-time decay of $F_{s}(t)$
is well fitted by a stretched exponential of the form
$\exp[-(t/t_r))^{\beta}]$, with relaxation time $t_r/\tau\cong
650$ and $\beta\cong 0.6$, once more confirming the heterogeneous
nature of the relaxation dynamics \cite{elmasri}; $t_r$ at
$\eta_2$ is beyond our simulation time.

Having established the {\em heterogeneity} of the dynamics in the
equilibrium smectic bulk phase, our next question concerns the
identification of the fast particles. Comparing the VHF at $t^*$
of Fig. 1(a) and (b) with a gaussian approximation with the same
MSD reveals that the fast-moving particles, which are responsible
for the non-gaussian character, have travelled over more than
$h/2$ during a time interval $t^*$. Therefore they are intimately
related to interlayer particles, which reside more than some
distance $\delta$ from the nearest smectic plane. In order to
define $\delta$ sensibly, we consider the variance of one period
of $\pi(z)$ as $\sigma^2\equiv \int_{-h/2}^{h/2}z^2 \pi(z)dz$,
which gives $\sigma/D=(0.6,0.3)$ for $\eta_1$ and $\eta_2$,
respectively, as indicated in Fig.1. We now set $\delta=k \sigma$
with $k=1,2,3$. The fraction $f_k$ of so-called interlayer
particles is then $f_1=(0.28,0.45)$, $f_2=(0.05,0.13)$ and
$f_3=(0.02,0.03)$ for our two state-points, showing, perhaps
surprisingly, that the denser state contains more interlayer
particles caused by the smaller $\sigma$. We also calculated the
pair distribution of the interlayer particles (not shown),
revealing a higher contact value than that of the bulk smectic
phase, suggesting substantial clustering of interlayer particles
despite their low concentration $f_k \eta$. Stringlike clusters
composed of $n=1,\cdots,10$ interlayer rods can indeed be
identified, as shown for $\delta=2\sigma$ by the size distribution
$P(n)$ in Fig. 3. Our
(rather stringent) cluster criterion is such that two interlayer
particles belong to the same cluster if the $z$ and $xy$ distance
are smaller than $h$ and $D$, respectively. At $\eta_{1}$, one
deduces from $P(n)$ that $\approx95\%$ of the clusters consist
mostly of 2 or 3 rods, while clusters of  more than 5 rods are
rare but do exist. The denser smectic phase has larger clusters,
which is again similar to supercooled liquids and glassy systems,
in which the cluster size increases with increasing cage-trapping
\cite{donati,berthier}. The fit $P(n)\propto \exp {(-\alpha n)}$
is accurate with $\alpha=(1.5,0.7)$ at $\eta_{1}$ and $\eta_{2}$,
respectively, from which the average cluster size follows as
$\langle n\rangle=(1-\exp(-\alpha))^{-1}$. These results depend,
however, strongly on $\delta$ as revealed by the inset of Fig. 3.

\begin{figure}[!ht]
\center
\includegraphics[width=0.8\textwidth]{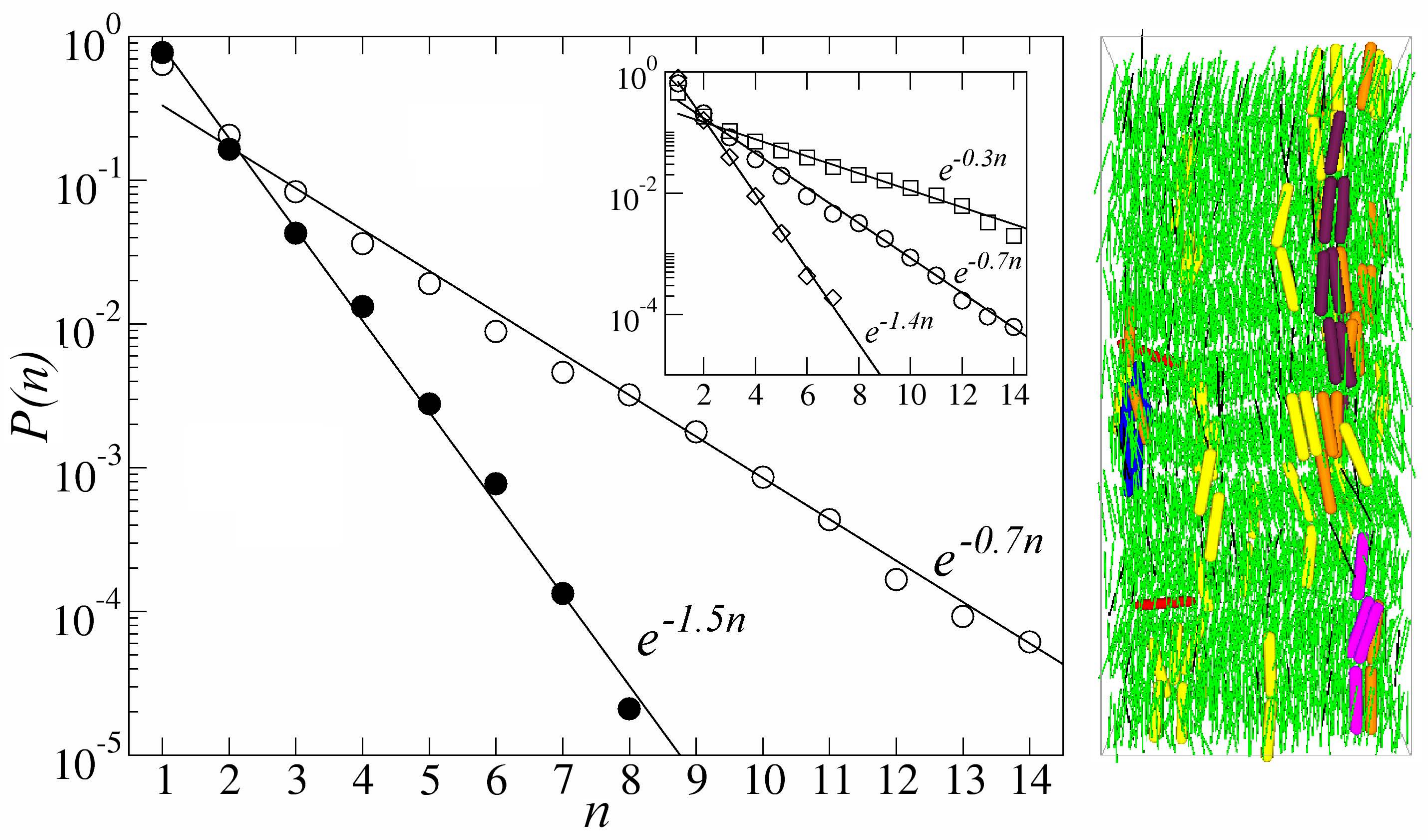}
\caption{Probability distribution $P(n)$ of the number $n$ of inter-layer rods in a stringlike cluster (with
$\delta=2\sigma$, see text), at $\eta_{1}$ (\CIRCLE) and $\eta_{2}$ (\Circle). The inset shows $P(n)$ at $\eta_2$
for $\delta=\sigma$ (\Square), $2\sigma$ (\Circle), and $3\sigma$ ($\Diamond$). The solid lines denote the fit $P(n)\propto\exp(-\alpha n)$ with $\alpha$ given in the figure. The snapshot of 3000 rods at $\eta_1$ shows
predominantly in-layer rods (green) and single interlayer rods (black), both with diameters reduced to $D/4$ for clarity. The thicker rods denote transverse ones (red), as well as clusters of 2 (yellow), 3 (orange), 4 (pink), 5 (blue), and $\geq 6$ (brown) rods. (color online).}
\end{figure}

With all these resemblances to glassy dynamics we now study the
dynamics of the layer-to-layer diffusion, for which Fig. 4b shows
some typical trajectories. Most of the rods "jump" fast compared
to the dwelling time within a smectic layer. Some rods diffuse to
the inter-layer spacing and return to their original layer (yellow
trajectory), others move from one layer to another several times
(blue), in some cases the dwelling time in the inter-layer spacing
is quite long (cyan), while double jumps can be also observed
(pink). These observations suggest a rather broad distribution of
jump-times $t_J$ to diffuse from one layer to the next, where
$t_J$ is defined as the time span between first and last "contact"
with the new and old layer, respectively, with contact established
if the rod is at a distance $\delta$ from the middle of a smectic
layer. At $\eta_1$ and for $\delta=2\sigma$, the distribution of
jump-times $\Pi(t_J)$  as obtained by averaging over many
trajectories is plotted in Fig. 4a, and shows that the
most-probable jump-time is $0.14 \tau$, the median is at
$t_J^*=0.2 \tau$, while the distribution extends over two decades
$0.01<t_J/\tau<1$. Having developed a sense of time, we can
characterize the trajectories further. We distinguish single and
multiple jumps, depending on the dwelling time in the next layer
being longer or shorter, respectively, than $t_J^*$ before another
jump is started. At $\eta_{1}$ we find that $\approx 5\%$ of the
total number of jumps is multiple, of which $\approx 83\%$ is a
double and the remaining fraction a triple jump. At $\eta_{2}$,
where $t_J^*=0.27 \tau$, the fraction of multiple jumps is
$0.6\%$. The inter-layer rods are generally oriented along
$\hat{n}$. Remarkably, however, at $\eta_{1}$ also transversely
oriented rods in between two smectic layers have occasionally been
observed \cite{vanroij,vanduijneveldt}, which diffuse either to a
new layer or return to the original one by rotating parallel or
anti-parallel to their original orientation.

With our clear evidence for heterogeneous dynamics and spatial
correlations between the fast-moving interlayer particles, it
seems natural to investigate dynamic cooperativity, i.e., whether
or not the stringlike clusters observed in static configurations
actually move collectively. This requires a cluster criterion that
not only involves spatial but also temporal proximity to identify
collectively moving rods. Two jumping rods $i$ and $j$ are
considered to move cooperatively if their arrival times $t^{(i)}$
and $t^{(j)}$ in their new layers (i.e. the first time at which
their distance to the middle of the new layer equals $\delta$)
satisfies $|t^{(i)}-t^{(j)}|<\Delta t$, while ${\bf r}_i(t^{(i)})$
and ${\bf r}_j(t^{(j)})$ satisfy our static spatial cluster
criterion. Using $\delta=2 \sigma$ and $\Delta t=\tau$ (i.e. long
enough for any jump to finish according to $\Pi(t_J)$), we find at
$\eta_1$ that the number of collective jumps is $\approx 25\%$ of
the total number of jumps, and involves mainly 2 ($\approx 79\%$)
or 3 ($\approx 17 \%$), and very rarely $\geq 4$ ($<4\%$) rods.
Interestingly, $\approx 42\%$ of the collective jumps involve rods
diffusing in opposite directions. These characteristic values are
insensitive to small modifications of our spatial cluster
criterion, while a smaller temporal interval of $\Delta t = t_J^*$
(the median jump time) reduces the fraction of collective jumps to
$\approx 9\%$. In other words, the motion is indeed strongly
cooperative at $\eta_1$. If the analogy with glassy systems 
would hold even further, one would expect {\em more} cooperativity  at
$\eta_{2}$. Surprisingly, perhaps, despite the larger static
clusters we find {\em less} collective motion at $\eta_2$.  For
instance, with $\delta=2\sigma$ and $\Delta t=1.2\tau$ (maximal
jump time) only $\approx 4\%$ of the jumps can now be regarded as
collective (involving essentially only 2 rods). We attribute this
reduction of cooperativity upon approaching crystallization to the
higher {\em permanent} barriers that reduce the probability of
static clusters to actually complete their attempted jumps. Such
permanent barriers do not exist in undercooled {\em liquids}
(where the dynamic cluster size grows upon approaching the glass
transition \cite{donati,berthier}), whereas they do exist in {\em
crystals}, where structural relaxation is virtually non-existent.

\begin{figure}[!ht]
\center
\includegraphics[width=0.8\textwidth]{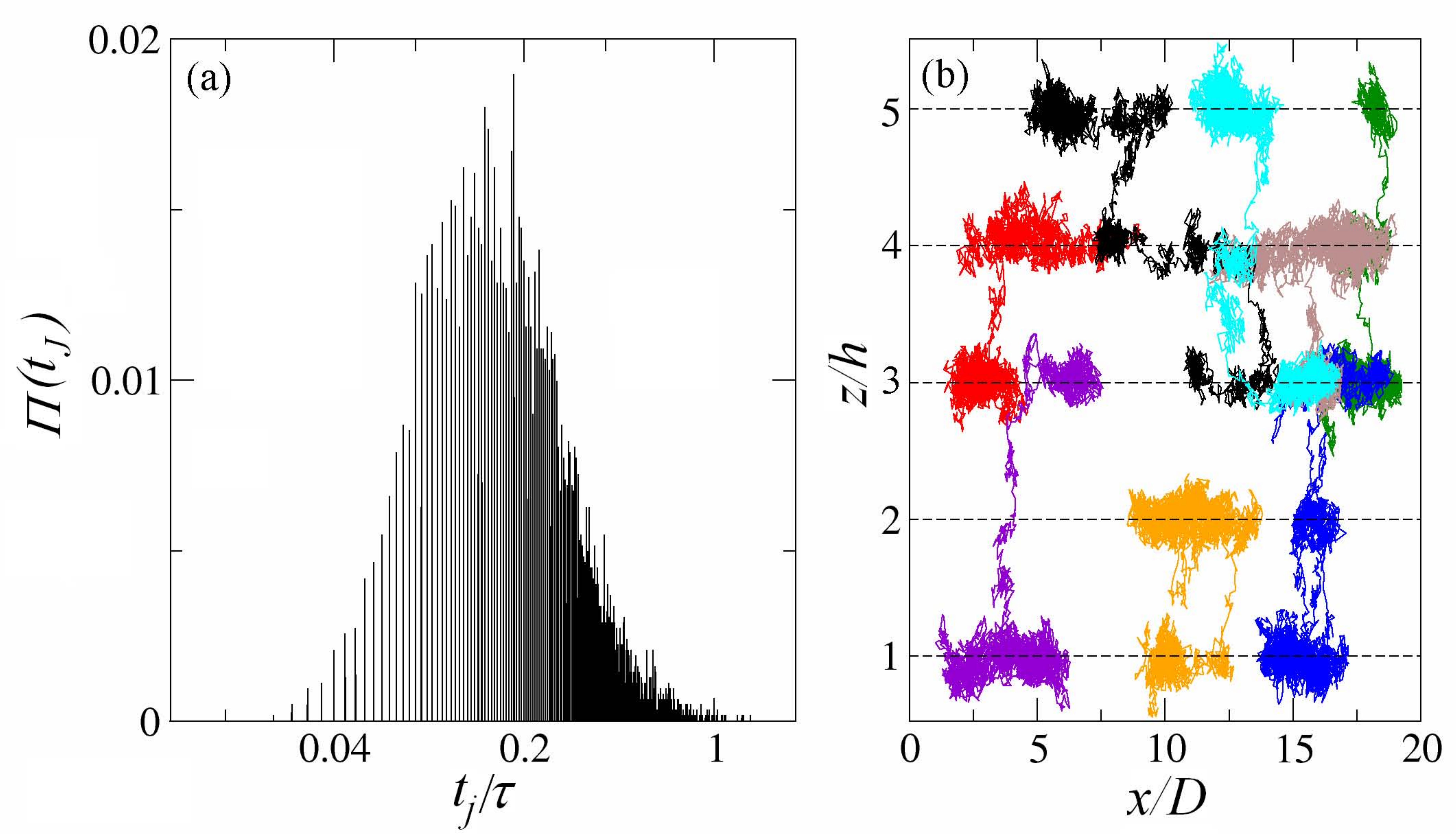}
\caption{ (a) Distribution $\Pi(t_J)$ of layer-to-layer jump times
$t_J$, based on $\delta=2\sigma$ at density $\eta_{1}$. (b)
Trajectories of jumping rods projected onto the \textit{xz} plane,
with the dashed lines representing the middle of the smectic
layers (color online).}
\end{figure}

In conclusion, for {\em equilibrium} hard rods in the smectic
phase we find non-gaussian and cooperative relaxation behavior
that is surprisingly similar to that of {\em non-equilibrium}
supercooled liquids. We attribute the non-gaussian diffusion to
the spatial inhomogeneous density of the smectic phase, and
connect this to dynamic heterogeneities and to static clusters of
inter-layer rods. Interestingly, we also find clear evidence for
cooperative motion of stringlike clusters. Our results might be
relevant for dynamics in other inhomogeneous liquids, e.g.,
confined fluids (micro and nanofluidic devices) \cite{mittal} and
columnar liquid crystals.

We thank P. van der Schoot and M. Bier for useful discussions. This work was financed by a NWO-VICI grant.

\end{document}